# Dipolar Magnetic Moment of the Bodies of the Solar System and the Hot Jupiters

## Hector Javier Durand-Manterola

Departamento de Ciencias Espaciales, Instituto de Geofísica,
Universidad Nacional Autónoma de México

hdurand_manterola@yahoo.com

Abstract

The planets magnetic field has been explained based on the dynamo theory, which presents as many difficulties in mathematical terms as well as in predictions. It proves to be extremely difficult to calculate the dipolar magnetic moment of the extrasolar planets using the dynamo theory.

The aim is to find an empirical relationship (justifying using first principles) between the planetary magnetic moment, the mass of the planet, its rotation period and the electrical conductivity of its most conductive layer. Then this is applied to Hot Jupiters.

Using all the magnetic planetary bodies of the solar system and tracing a graph of the dipolar magnetic moment versus body mass parameter, the rotation period and electrical conductivity of the internal conductive layer is obtained. An empirical, functional relation was constructed, which was adjusted to a power law curve in order to fit the data. Once this empirical relation has been defined, it is theoretically justified and applied to the calculation of the dipolar magnetic moment of the extra solar planets known as Hot Jupiters.

Almost all data calculated is interpolated, bestowing confidence in terms of their validity. The value for the dipolar magnetic moment, obtained for the exoplanet Osiris (HD209458b), helps understand the way in which the atmosphere of a planet with an intense magnetic field can be eroded by stellar wind. The relationship observed also helps understand why Venus and Mars do not present any magnetic field.

## 1 Introduction

Several credited theories exist, which attempt to provide an explanation for the magnetic field of a planet. The first, proposed by William Gilbert in 1600, in his book *De Magnete* was that of permanent magnetization; or in other words, a planet body behaves as a permanent magnet. Over time, other explanations have emerged, such as electric charge in rotation in the planetary body (Sutherland, 1900; Parkinson, 1983), the free fall of electric currents (Lamb, 1932; Rikitake, 1966), giro-magnetic effect (Barnett, 1933), thermo-electrical effects (Elsasser, 1939), Hall's effect (Vestine, 1954), differential rotation (Inglis, 1955), electromagnetic induction caused by magnetic storms (Chatterjee, 1956), etc. Currently, the most accepted explanation is the dynamo model. In spite of the fact that this represents the best candidate for inducing the magnetic field of a planet, it presents some problems. For example, the theory predicts that the axis of the dipole cannot be parallel to the axis of rotation (Ferraro and Plumpton, 1966, pp 2; Alexeff, 1989) and yet Saturn has these two axes parallel to each other (de Pater and Lissauer, 2001, pp 264). In short, the exact mechanism for generating the magnetic field of the planets is not known with certainty. And the various theories which have been developed, based on the dynamo model, have not managed to explain all aspects of this phenomenon.

We know from the Ampere-Maxwell law that a magnetic field can be generated in two ways; either by electric currents or by temporal variations in the electric field. In the interior of a planet there may be free charges which generate electrical fields, which, when varied, (because of the separation and recombination of charges) are able to generate magnetic fields. However, because of the neutral, global character of the material that forms planets, we may assume that the separations of charge that occur will not be very great and because of this, the electric and magnetic fields produced will not be very intense. In the case of electric currents, these can produce an intense magnetic field, so that in the case of the magnetic planetary fields it is very probable that these are generated by currents in the interior of the body. These currents may constitute molecular currents, like those that generate permanent magnetism. However, given the temperatures that prevail in the interior of the planets, temperatures exceeding Curie's temperature point; this magnetism could only exist in the external layers of the planet, where temperatures are lower. We do not know what level of magnetism exists in the outer layer of the Mercury, but in the case of the Earth, the magnetic intensity of the planet by far exceeds the magnetism of the rocks of the outer layer and in the case of the gaseous giants, the upper layers are conformed of gas, thus impeding the presence of any permanent magnetism. All these arguments lead us to reject the idea of permanent magnetism or that molecular currents should be a source of planetary magnetism; and thus the only remaining possibility is that they represent macroscopic currents. For this to be possible, a conducting material must be present within the body of the planet where the currents are produced. If this is the case, the magnetic moment of the planet should therefore depend on its conductivity and on the quantity of conducting material present, which may also depend on the total mass of the planet. If the mechanism that generates the magnetic field is of the dynamo type, then the magnetic moment will also depend on the speed at which the planet rotates, or, in other words, its rotation period.

In this study, these three parameters are correlated: mass m, rotation period P, and electric conductivity σ for all the magnetic planets of the solar system, together with their dipolar magnetic moment M, in order to obtain an empirical equation, allowing the calculation of the magnetic moment for the extra solar planets. Section 2 describes the relationship that exists between the magnetic moment and the mass of the planet; section 3 describes the relationship between the magnetic moment and the rotation period of the planet.; section 4 describes the relationship that exists between the magnetic moment and electric conductivity; section 5 describes the relationship of the magnetic moment with a function of the three previous parameters; section 6 provides a rough outline of the theoretical justification for the empirical formula. And in section 7 the empirical formula is employed to calculate the magnetic moment of certain extra solar planets; those denominated as the "Hot Jupiters"

## 2 The functional relationship between magnetic moment and mass

Seven magnetic planets are known to exist in the solar system: Jupiter, Saturn, Uranus, Neptune, Earth, Mercury and Ganymede.

From the measurements of the magnetic moment and the mass of these seven planetary objects (see Table I), a curve may be fitted to the data as can be seen in figure 1. The relationship between the magnetic moment and the mass is a power law:

$$M = 2 \times 10^{-21} m^{1.7365} \quad (1)$$

Hence, increasing planet mass increases magnetic field. This relationship between the data has a correlation coefficient of 0.9804, reinforcing the idea that a physical relationship exists between these two variables. If it is assumed, as with the dynamo theory, that the magnetic field is produced by electric currents in a conducting layer, then relationship (1) indicates that if planet mass increases, the electric current that generates the magnetic field also increases. This is coherent, as mass increases, we also expect the amount of conductor material to increase, so the electric current will be more intense and consequently the magnetic field will be stronger. However, the generation of the magnetic field does not only depend on the mass. If we apply equation (1) to the case of the planet Venus, according to its mass it should have a magnetic moment similar to that of the earth ($3.953 \times 10^{22}$ Amp-m$^2$; Cox, 2000). However in reality, its magnetic moment is either non-existent or less than $5 \times 10^{17}$ Amp-m$^2$ (Cox, 2000). This makes us conclude that even though planetary mass may be significant; it is not the only factor intervening in the generation of a magnetic field.

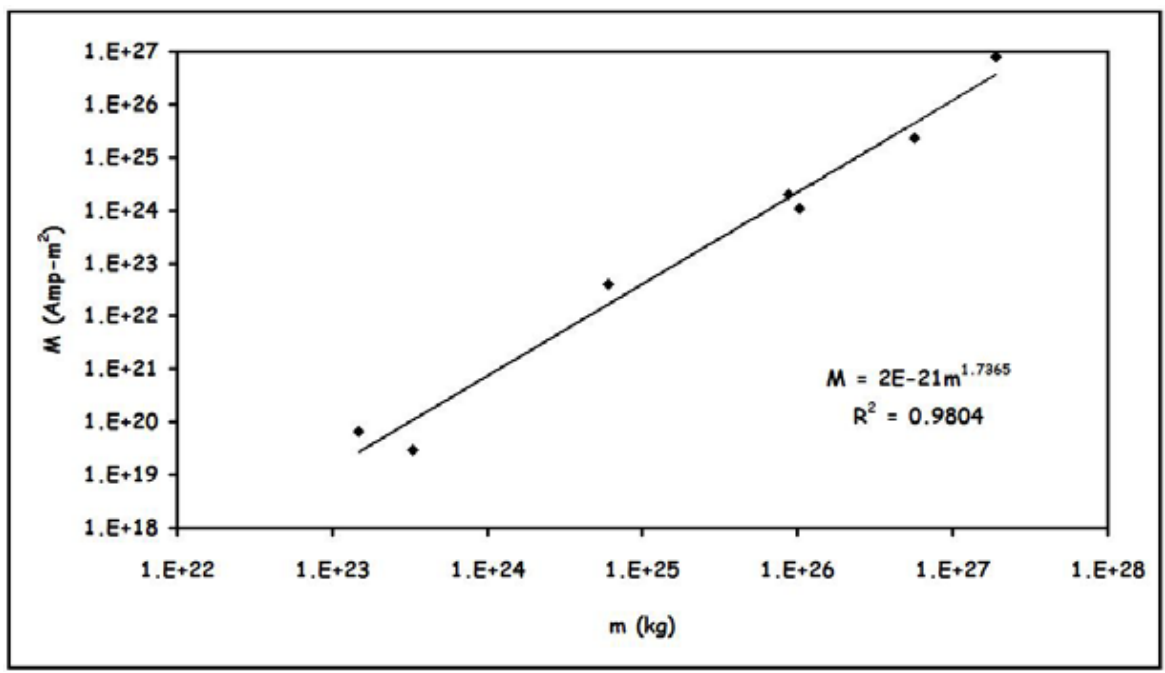

*Figure 1 Functional relationship between the planetary dipolar magnetic moment and the mass for the seven magnetic planetary bodies of the solar system.*

Table I
Data of the Solar System Magnetic Planets

| Planet | Mass m (x$10^{24}$ kg) | Mass Error (kg) | Period P (s) | P. Error (s) | Dipolar Magnetic Moment M (Amp-m$^2$) | M. Error (Amp-m$^2$) |
|---|---|---|---|---|---|---|
| Mercury * | 0.33022 | ± 5x$10^{18}$ | 5067031 | ± 5 | 4x$10^{19}$ | ± 5x$10^{18}$ |
| Earth * | 5.9742 | ± 5x$10^{19}$ | 86164.1003 | ± 5e-4 | 7.84x$10^{22}$ | ± 5x$10^{19}$ |
| Jupiter * | 1898.7 | ± 5x$10^{22}$ | 35729.8 | ± 0.5 | 1.55x$10^{27}$ | ± 5x$10^{24}$ |
| Saturn * | 568.51 | ± 5x$10^{21}$ | 38362.4 | ± 0.5 | 4.6x$10^{25}$ | ± 5x$10^{23}$ |
| Uranus * | 86.849 | ± 5x$10^{20}$ | 62063.7 | ± 0.5 | 3.9x$10^{24}$ | ± 5x$10^{22}$ |
| Neptune * | 102.44 | ± 5x$10^{21}$ | 57996 | ± 0.5 | 2.2x$10^{24}$ | ± 5x$10^{22}$ |
| Ganymede + | 0.148186 | ± 5x$10^{17}$ | 618153.3757 | ± 5x$10^{-4}$ | 1.32x$10^{20}$ | ± 5x$10^{17}$ |

\* Tholen et al., 2000
+ Kivelson et al., 2002

## 3 Functional relationship between the magnetic moment and the rotation period

Once again, considering the data but now referring to the dipolar magnetic moment and the rotation period for the seven magnetic planets (see Table I), we can observe in figure 2 that in the case of these planets, the functional relationship between the magnetic moment and the rotation period represents an exponential function, meaning, that when the inverse of the rotation period is increased or the rotation period is decreased; the magnetic moment increases and can be expressed as:

$$M = 4 \times 10^{19} \exp\left(\frac{579696}{P}\right) \qquad (2)$$

This relationship between the data has a correlation coefficient of 0.9742, reinforcing the idea that a physical relationship exists between these two variables. However, the generation of the magnetic field does not only depend on the rotation period. If we were to apply equation (2) to the case of the planet Mars, its rotation should induce a magnetic moment similar to that of the Earth (3.953x$10^{22}$ Amp-m$^2$; Cox, 2000). However, in reality, its dipolar moment is non-existent or less than 5x$10^{18}$ Amp-m$^2$.(Cox, 2000) Thus, we can conclude that although the rotation period is important; it is not the only factor which intervenes in the generation of the magnetic field.

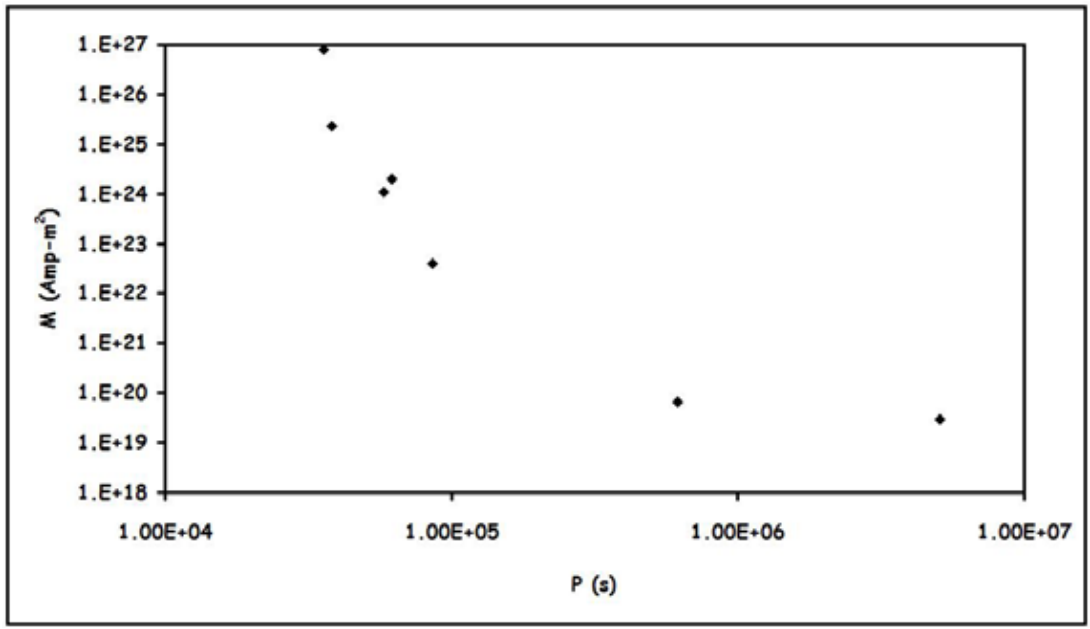

*Figure 2 Functional relationship between the dipolar magnetic moment and the rotation period for the seven magnetic planetary bodies of the solar system.*

**4 Functional relationship of M with σ**

Contrary to mass and rotation periods, which are both parameters that can be measured, the value of conductivity in the conducting layer is not easy to calculate. From models of the interior of the planet Earth, iron-nickel (Fe-Ni) alloy is the most abundant component in the Earth's core (Lin et al., 2002) forming a conductive layer with a mean electric conductivity of $1.2 \pm 0.2 \times 10^5$ S/m, between $1.02\pm0.005 \times 10^5$ S/m (Iron conductivity) and $1.43\pm0.005 \times 10^5$ S/m (nickel conductivity) (Kittel, 2004). Models for the interior of Mercury also suggest that the interior is made of iron-nickel (Hamblin and Christiansen, 1990) which is responsible for Mercury's magnetic field. Some contemporary geodynamic models say that shear motions between the core and the base of the Earth's mantle drives the geodynamo (Labrosse et al., 2007, and Labrosse et al., 2003). If this is the case, the composition of this layer is uncertain but most probably enriched by iron. The core conductivity upper limit is probable and so will be used here.

In the case of the gaseous giants, it has been suggested that the conductive layer is formed from metallic hydrogen, with a conductivity of $2 \pm 0.5 \times 10^5$ S/m (Shvets, 2007).

In the case of Ganymede, Kivelson et al. (2002) suggest that salt water forms the conductive layer. As we do not know the concentration of these salts, it would be risky to calculate the conductivity of the Ganymedean Ocean. The maximum conductivity of the Earth's seawater is 6.5683 S/m (Kennish, 2001), so this may provide a first approximation for Ganymede. However, the absence of a magnetic field in Europa or Callisto (Gurnett et al., 1997), most of all in Europa, where there are indications that a sub-superficial ocean exists throws doubts on the possibility that an ocean of this type would be capable of generating a magnetic field, such as that which is calculated for Ganymede. For this reason and others mentioned later in this study, a conductivity of 1.2 ± 0.2 x$10^5$ S/m was attributed to Ganymede, assuming it has an iron nucleus.

With only two conductivity values for the seven planets it would be pointless to try and calculate the functional relationship between σ and M, however, in figure 3, we can see that all the bodies with less conductivity (those of iron-nickel) have magnetic moments which fall below those with layers of metallic hydrogen (greater conductivity).

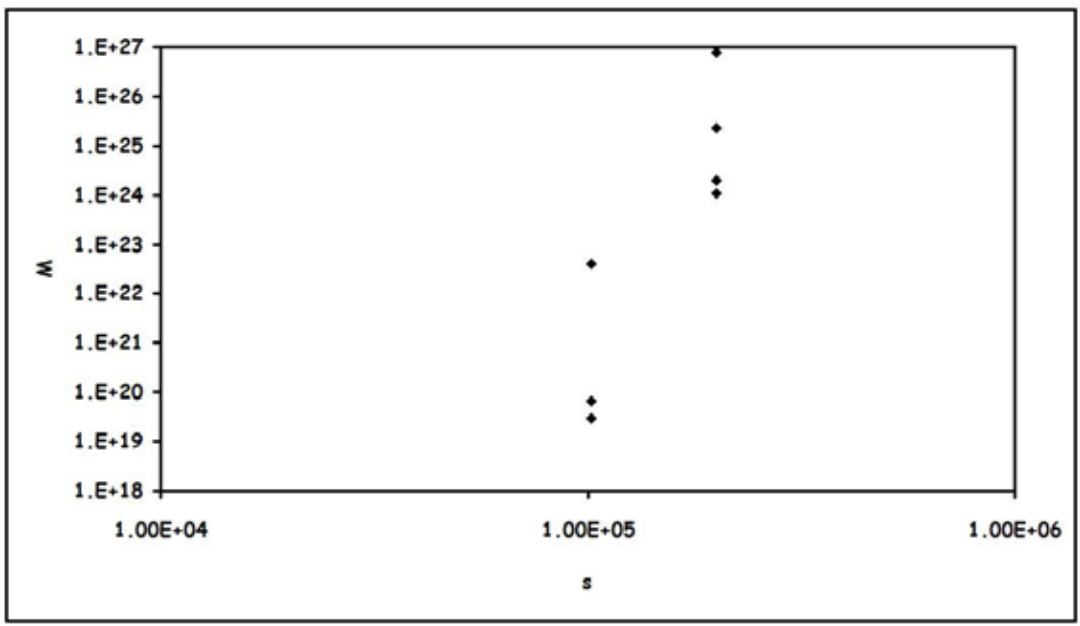

*Figure 3 Relationship between the dipolar magnetic moment and electrical conductivity.*

### 5 Functional relationship of M with mσ/P

Finally, putting together all three factors under a single parameter mσ/P; the values of the seven planets constitute a power law curve which we can observe in figure 4

$$M = 1 \times 10^{-5} \left[ \frac{m\sigma}{P} \right]^{1.106} \quad (3)$$

In order to determine the variation interval from the inclination of the curve (exponent of equation 3), various fittings for the data were carried out, in each case removing one of the planets and calculating the value for this group of data. In Table II, the values are presented for each case. The average and error were calculated for extreme values.

In the past some authors proposed a potential relationship between the magnetic moment (M) and the angular momentum (L), the so-called magnetic Bode's law (Blackett, 1947, Kennel, 1973; Hill and Michel, 1975; Dessler, 1976; Siscoe, 1978; Russell, 1978). Because no physical justification existed some other authors question it (Cain et al, 1995). In the next section we derive equation (3) from the first principles.

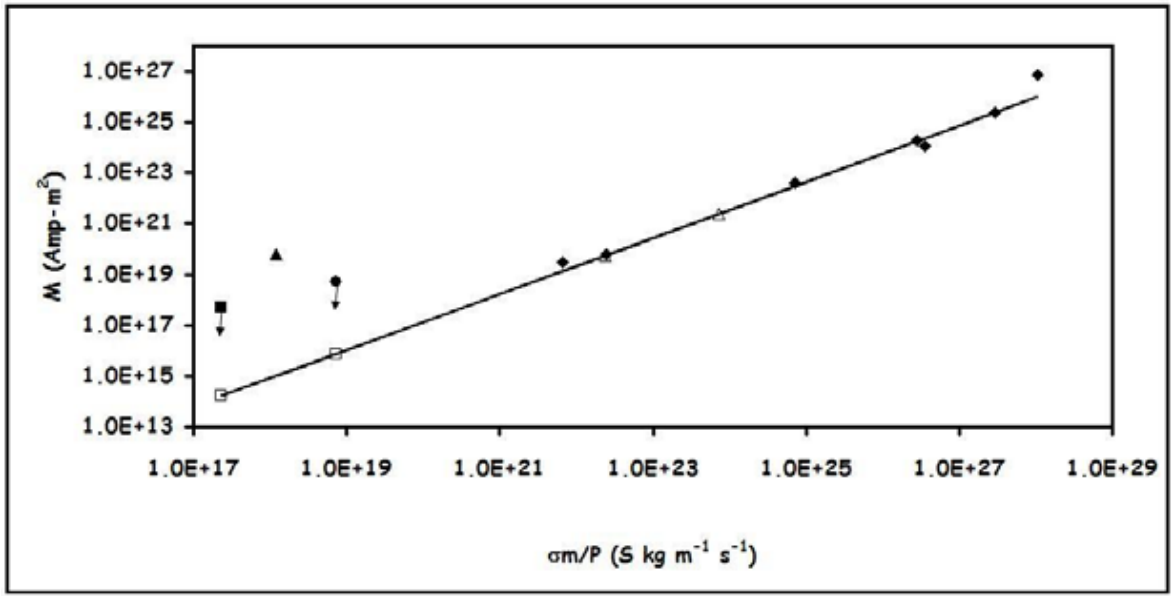

Figure 4 Planetary magnetic moment as a function of the parameter σm/P, where σ represents the conductivity of the conductor layer of the planet, m its mass and P the rotation period of the body. The diamonds represent the 7 magnetic bodies of the solar system. The straight line represents the graph for equation (3). The filled in triangle is Ganymede only if its conductivity were the same as that of sea water. The filled in circle and square represent the maximum values for the magnetic moment of Mars and Venus, respectively, graphed against the parameter σm/P, assuming a conductivity of 1 S/m which is the conductivity of silicates (perovskite) The empty triangles and squares

*represent the predicted values for equation (3), in the case that Venus and Mars have either a conductivity of iron-nickel (triangles) or silicates (squares).*

Table II
Variation of the exponent of equation (3)
Eliminating one planet of the seven. $R^2$ is the
Correlation coeficient of data

|  | Exponent | Error of the Average | $R^2$ |
|---|---|---|---|
| All | 1.1218 |  | 0.9814 |
|  |  |  |  |
| Without Mercury | 1.1685 |  | 0.975 |
| Without Earth | 1.1212 |  | 0.9813 |
| Without Jupiter | 1.0435 |  | 0.9959 |
| Without Saturn | 1.1388 |  | 0.9805 |
| Without Uranus | 1.1267 |  | 0.9814 |
| Without Neptune | 1.1418 |  | 0.9878 |
| Without Ganymede | 1.1143 |  | 0.9739 |
| ------ | ------ | ------ | ------ |
| Larger | 1.1685 |  |  |
| Minor | 1.0435 |  |  |
| Average | 1.10 | $\pm 0.13$ |  |

## 6 Theoretical justifications for formula (3)

The scalar magnetic potential for a dipole seen from a distance r is (Reitz y Milford, 1981, pp 178):

$$U(r) = \frac{Mr\cos(\theta)}{4\pi r^3} \quad (4)$$

Where θ is the angle between the vector **M** and the vector **r**.
On the other hand, the scalar magnetic potential of any circuit seen from afar is (Reitz and Milford, 1981, pp 179):

$$U(r) = -\frac{I\Omega}{4\pi} \quad (5)$$

Where I is the current in the circuit and Ω is the solid angle from which the circuit can be seen from a distance r.

Near to the planet, the magnetic field is seen as a magnetic dipole, and because of this (4) and (5) can be made equal in order to obtain M. We also substitute the electric current for s$\varepsilon$, where s represents electric conductance and $\varepsilon$ represents the electromotive force. And, if we are placed at the magnetic equator of the planet, where cos ($\theta$) = 1 and r = R, R representing the radius of the planet, then, $4\pi R^3$ = V represents the volume of the planet and s/$2\pi R$ = $\sigma$ represents electrical conductivity. Thus, the dipolar moment can be expressed as:

$$M = -\frac{\sigma \varepsilon \Omega}{2} V \qquad (6)$$

From Faraday's law:

$$\varepsilon = -\frac{d\Phi_B}{dt} \qquad (7)$$

And by the chain rule:

$$\varepsilon = -\frac{d\Phi_B}{dt} = -\frac{d\Phi_B}{d\varphi}\frac{d\varphi}{dt} = -\frac{d\Phi_B}{d\varphi}\omega \qquad (8)$$

And from equations (6) and (8) we have:

$$M = \frac{\Omega}{2}\frac{d\Phi_B}{d\varphi}\sigma V \omega \qquad (9)$$

But $\omega$ = $2\pi$/P. We substitute this and multiply the numerator and the denominator by $\rho$ the average density of the planet. On the other hand $\rho V$ = m. Then equation (9) is expressed as:

$$M = \frac{\pi \Omega}{\rho}\frac{d\Phi_B}{d\varphi}\frac{\sigma m}{P} \qquad (10)$$

That is to say

$$M \propto \frac{\sigma m}{P} \qquad (11)$$

In equation (3), the exponent is different from unity, but close to it.

# 7 Magnetic fields of the "Hot Jupiters"

If relationship (3) represents a general formula for planets, then it will serve for predicting the magnetic field of extrasolar planets known as "Hot Jupiters". The reason "Hot Jupiters" were chosen from among the other exoplanets is because they are located at short distance from the star, we can assume that their rotation has been trapped and therefore it is the same as the period it takes to travel around the star. Once its rotation period and mass are known, assuming that these are gaseous planets, we can use the conductivity of metallic hydrogen in order to calculate the parameter σm/P. Using formula (3), the magnetic moment can be calculated. Table III shows the calculated values for 31 Hot Jupiters. Inserted into the same table were Rss values of the distance from planet to the sub stellar point of the magnetopause calculated with the formula:

$$Rss = \left[\frac{M^2}{4\pi\rho V^2}\right]^{1/6} \quad (12)$$

# 8 Discussion and Conclusions

In the case of Mars and Venus, because they are terrestrial planets, one would expect them to have an iron-nickel nucleus. If we take the conductivity of iron for Venus and Mars and use relationship (3) to calculate M, it gives us a greater magnetic moment than that of Mercury, greater than the upper limits measured for either planet (figure 4, empty triangles). It is known that in the past Mars had a magnetic field (Acuña et al, 2001) and for some reason it disappeared. It is possible that great impacts, such as those which produced Argyre and Hellas could eliminate the dynamo which generated the magnetic field. That impact could have injected material with a low conductivity (i.e. silicates) into the planet's nucleus.

In figure 4 (filled square and circle), we see the maximum estimate for the magnetic moment of both these planets, which has been calculated based on the observations. The arrows indicate that the real value of the magnetic moment must be lower than these values. In the same figure (empty squares) we can observe what the magnetic moment for both planets would be if we reduce the conductivity, assigning them that of silicates (perovskita) and applying equation (3). The idea here is that even though metals exist in the nucleus, these do not necessarily form a continuous mass, but are instead mixed with a material of poor conductivity (silicates) and that these are unable to generate generalized currents for the entire nucleus. In the case of Venus another possibility exists. The Earth's magnetic field has been inverted countless times, as indicated in studies of oceanic rocks. However, we do not know what happens when these inversions occur. We do not know whether the magnetic field diminishes until it disappears and then grows in an inverted way, or if it simply rotates (Merrill and McFaden, 1999). If it were the former, how long does this period of field zero endure? It is known from the oceanic rocks that this moment is geologically rapid. However, one thousand years is rapid in geological terms. Is it possible that Venus is currently going through one of these moments? Some estimates of the time of an inversion lie between 1000 and 8000 years (Merrill and McFaden, 1999). The slow retrograde rotation of Venus would anyways lead to a theoretical value of 0.33% of the Earths magnetic moment. According to some contemporary geodynamic models (Labrosse et al., 2007, and Labrosse et al., 2003) one may also conceive the absence of a magnetic dynamo as the cessation of intrinsic shear motion between the inner mantle and the nucleus induced by a catastrophic event.

In the same figure, it can be seen that the relationship between the parameter $m\sigma/P$ and M for Ganymede, does not coincide with the relationship found for other planets using the conductivity of sea water (5S/m) (full triangle). On the other hand, using the conductivity of iron-nickel, it would fall within the relationship. This appears to indicate that within the interior of Ganymede, there is a conducting, metallic nucleus.

Taking the average density of Ganymede ($\rho$ =1,406 kg/m$^3$) and its radius (R = 2.634x10$^6$ m) and assuming an interior model of 2 layers; a covering of ice ($\rho_m$ = 1000 kg/m$^3$) and a nucleus of iron-nickel ($\rho_n$ = 11,000 kg/m$^3$), we can calculate the radius of the nucleus applying the following formula

$$r = \left[ \frac{\rho - \rho_m}{\rho_n - \rho_m} \right] R \qquad (12)$$

This radius would be 107 km; that is to say, a very small nucleus. It may be assumed that among the components of the satellites of the Giant planets, metals are present but in the case of the majority, the metal is mixed with non-conducting compounds, without forming a continuous mass and because of this, they cannot create strong electric currents. This is the reason that bodies such as Titan, Europa and Callisto do not have a magnetic field. However, in the case of Ganymede this being the largest satellite, it is possible that the differentiation of the body has been total and that the metal has been separated, forming an iron nucleus.

In Table 1, we can observe an interesting result, when all the magnetic planets are correlated, or if we remove one that is not Jupiter, the exponent will always exceed 1.1. However, in the moment we remove Jupiter, the exponent reduces to almost unity and the correlation coefficient is almost one. This tells us that Jupiter may have an unusually intense magnetic field, either because the planet is anomalous, or because it is passing through a brief interval of greater intensity than usual.

The benefit that the values for the magnetic moment can provide for the exoplanets is that with these, together with the radius of the planet (if known); the magnetic intensity of the surface of the planet can be calculated. Similarly, if the value of the magnetic field at the surface is known, then the distance of the sub stellar point of the magnetopause from the body may be found (see Table III).

Schneider et al. (1998) predicted that exoplanets of the Hot Jupiter type would have a cometary exosphere. Schneiter et al. (2007) made simulations for the interaction of stellar wind with the expanded atmosphere of the Hot Júpiter HD209458b ("Osiris"), obtaining this type of cometary exosphere. They concluded that the existence of a great stellar cometary wake suggests that Osiris does not possess a global magnetic field.

In this work, the magnetic moment calculated for Osiris is equal to $1.26 \times 10^{26}$ Amp-m$^2$, which is equivalent to 5.5 times the magnetic moment of Saturn and 1/6 that of Jupiter. Sánchez-La Vega (2004) using a completely different method came to the same conclusion, that Osiris has a magnetic moment that is intermediate between that of Saturn and of Jupiter.

This means an intense magnetic field, but does not necessarily contradict the results of Schneiter et al. (2007). Osiris is located at 0.045 U.A. from the star and for this reason its atmosphere is evaporating. The neutral gases escape from the magnetosphere without problem. The particles ionised by the EUV and X-rays radiation out of the plasmasphere are convected to the front of the magnetosphere and expelled to the stellar wind. It is also possible that an interaction between Osiris and its star (Preusse et al., 2006) permit the escape of gas from the cusps of the planet magnetosphere.

From this study the following conclusions can be made:
- Using known data from the solar system, it is possible to construct a function which permits us to calculate the magnetic moment for the Hot Jupiters.
- Theoretical justification of the above function was obtained.

- The values calculated for magnetic moment are mostly interpolated.
- Once the values for magnetic moments are known; it is possible to calculate the sub stellar point of the magnetopause and subsequently the size of the magnetosphere.
- The value for magnetic moment calculated for the planet Osiris permits us to understand how a body with a magnetic field can simultaneously be eroded by solar wind.
- The same function permits us to understand the very weak magnetic moment, manifested by Mars and Venus, assuming that the metallic nucleus underwent a mixing of silicates from the mantle, leading to catastrophic events.

Table III

Magnetic Moment M and sub stellar distance Rss of the magnetopause calculated for 31 Hot Jupiters

|   | Planet | Mass m (kg) | Orbital period P (s) | mσ/P (kg S/ms) | M (Amp-m$^2$) | Rss (km) |
|---|---|---|---|---|---|---|
| 1 | OGLE-TR-56 b | 2.7538E+27 | 1.0471E+05 | 5.2599E+27 | 9.3346E+26 | 4.08e6 |
| 2 | OGLE-TR-113 b | 2.5639E+27 | 1.2377E+05 | 4.1431E+27 | 7.169E+26 | 3.76e6 |
| 3 | GJ 436 b | 1.2724E+26 | 2.2843E+05 | 1.1141E+26 | 1.314E+25 | 1.06e6 |
| 4 | OGLE-TR-132 b | 2.26E+27 | 1.4600E+05 | 3.0958E+27 | 5.1939E+26 | 3.72e6 |
| 5 | HD 63454 b | 7.2169E+26 | 2.4346E+05 | 5.9286E+26 | 8.3479E+25 | 2.13e6 |
| 6 | HD 73256 b | 3.5515E+27 | 2.2020E+05 | 3.2257E+27 | 5.4354E+26 | 4.02e6 |
| 7 | 55 CnC e | 8.5463E+25 | 2.4278E+05 | 7.0402E+25 | 7.9091E+24 | 9.89e5 |
| 8 | TrES-1 | 1.1585E+27 | 2.6180E+05 | 8.8503E+26 | 1.3003E+26 | 2.54e6 |
| 9 | HD 83443 b | 7.7866E+26 | 2.5796E+05 | 6.0371E+26 | 8.5171E+25 | 2.22e6 |
| 10 | HD 179949 b | 1.8612E+27 | 2.6719E+05 | 1.3931E+27 | 2.1476E+26 | 3.02e6 |
| 11 | HD 46375 b | 4.7289E+26 | 2.6127E+05 | 3.6199E+26 | 4.8374E+25 | 1.85e6 |
| 12 | OGLE-TR-10 b | 1.0825E+27 | 2.6795E+05 | 8.08E+26 | 1.1757E+26 | 2.51e6 |
| 13 | HD 187123 b | 9.8757E+26 | 2.6758E+05 | 7.3815E+26 | 1.0638E+26 | 2.43e6 |
| 14 | HD 330075 b | 1.4434E+27 | 2.9108E+05 | 9.9173E+26 | 1.4747E+26 | 2.73e6 |
| 15 | HD 2638 b | 9.116E+26 | 2.9758E+05 | 6.1268E+26 | 8.6571E+25 | 2.31e6 |
| 16 | HD 209458 b | 1.3104E+27 | 3.0454E+05 | 8.606E+26 | 1.2606E+26 | 2.63e6 |
| 17 | BD -10 3166 b | 9.116E+26 | 3.0136E+05 | 6.0499E+26 | 8.537E+25 | 2.33e6 |
| 18 | HD 75289 b | 7.9765E+26 | 3.0326E+05 | 5.2605E+26 | 7.3138E+25 | 2.21e6 |
| 19 | HD 88133 b | 4.1782E+26 | 2.9462E+05 | 2.8363E+26 | 3.6935E+25 | 1.77e6 |
| 20 | OGLE-TR-111 b | 1.0066E+27 | 3.4685E+05 | 5.804E+26 | 8.1542E+25 | 2.31e6 |
| 21 | HD 76700 b | 3.7414E+26 | 3.4309E+05 | 2.181E+26 | 2.7621E+25 | 1.63e6 |
| 22 | Tau-Boo | 7.8436E+27 | 2.8629E+05 | 5.4795E+27 | 9.7666E+26 | 5.39e6 |
| 23 | 51 Peg b | 8.8881E+26 | 3.6554E+05 | 4.863E+26 | 6.7052E+25 | 2.24e6 |
| 24 | HD 49674 b | 2.279E+26 | 4.2714E+05 | 1.0671E+26 | 1.2528E+25 | 1.32e6 |
| 25 | ups-And b | 1.3104E+27 | 3.9892E+05 | 6.57E+26 | 9.3523E+25 | 2.61e6 |
| 26 | HD 168746 b | 4.3681E+26 | 5.5322E+05 | 1.5792E+26 | 1.9326E+25 | 1.59e6 |
| 27 | HD 217107 b | 2.4309E+27 | 6.1576E+05 | 7.8957E+26 | 1.1461E+26 | 2.95e6 |
| 28 | HD 68988 b  | 3.6084E+27 | 5.4225E+05 | 1.3309E+27 | 2.0418E+26 | 3.59e6 |
| 29 | HD 162020 b | 2.6114E+28 | 7.2820E+05 | 7.1721E+27 | 1.3153E+27 | 6.72e6 |
| 30 | HD 130322 b | 2.0511E+27 | 9.2655E+05 | 4.4274E+26 | 6.0441E+25 | 2.57e6 |
| 31 | HD 160691 d | 7.9765E+25 | 8.2512E+05 | 1.9334E+25 | 1.894E+24  | 8.16e5 |